\documentclass[twocolumn,prl,superscriptaddress,amsmath]{revtex4}  
\usepackage{graphicx}
\newcommand{\avk}{\langle k \rangle}

\begin{document}

\title{Non-equilibrium phase transition in negotiation dynamics}

\author{Andrea Baronchelli} 
\affiliation{Dipartimento di Fisica, Universit\`a
  ``La Sapienza'' and SMC-INFM, P.le A. Moro 2, 00185 Roma (Italy)}
\affiliation{Departament de F\'\i sica i Enginyeria Nuclear, Universitat
  Polit\`ecnica de Catalunya, Campus Nord B4, 08034 Barcelona (Spain)}
\author{Luca Dall'Asta} 
\affiliation{LPT, CNRS (UMR 8627) and Univ Paris-Sud, Orsay, F-91405  (France)} 
\affiliation{Abdus Salam International Center for Theoretical Physics, Strada
  Costiera 11, 34014, Trieste (Italy)} 
\author{Alain Barrat}
\affiliation{LPT, CNRS (UMR 8627) and Univ Paris-Sud, Orsay, F-91405  (France)} 
\affiliation{Complex Networks Lagrange Laboratory, ISI Foundation, 
Turin (Italy)}
\author{Vittorio Loreto} 
\affiliation{Dipartimento di Fisica, Universit\`a
``La Sapienza'' and SMC-INFM, P.le A. Moro 2, 00185 ROMA, (Italy)}

\date{\today}

\begin{abstract}
We introduce a model of negotiation dynamics whose aim is that of
mimicking the mechanisms leading to opinion and convention formation
in a population of individuals. The negotiation process, as opposed to
``herding-like'' or ``bounded confidence'' driven processes, is based
on a microscopic dynamics where memory and feedback play a central
role. Our model displays a non-equilibrium phase transition from an
absorbing state in which all agents reach a consensus to an active
stationary state characterized either by polarization or fragmentation
in clusters of agents with different opinions. We show the exystence
of at least two different universality classes, one for the case with 
two possible opinions and one for the case with an unlimited number
of opinions. The phase transition is studied analytically and
numerically for various topologies of the agents' interaction
network. In both cases the universality classes do not seem to depend
on the specific interaction topology, the only relevant feature being
the total number of different opinions ever present in the system.

\end{abstract}

\maketitle

Statistical physics has recently proved to be a powerful framework to address 
issues related to the characterization of the collective social behavior 
of individuals, such as culture dissemination, the spreading of linguistic 
conventions, and the dynamics of opinion formation~\cite{sociophysics}.

According to the ``herding behavior'' described in
sociology~\cite{herd}, processes of opinion formation are usually
modeled as simple collective dynamics in which the agents update their
opinions following local majority~\cite{majority} or imitation
rules~\cite{voter}. Starting from random initial conditions, the
system self-organizes through an ordering process eventually leading
to the emergence of a global consensus, in which all agents share the same opinion. 
In analogy with kinetic Ising models and contact processes~\cite{odor},
the presence of noise can induce non-equilibrium phase transitions
from the {\em consensus} state to disordered configurations, in which
more than one opinion is present.\\ The principle of ``bounded confidence''~\cite{deffuant,krause}, 
on the other hand, consists in
enabling interactions only between agents that share already some
cultural features (defined as discrete objects)~\cite{axelrod} or with
not too different opinions (in a continuous
space)~\cite{deffuant,bkr}. 
By tuning some threshold parameter, transitions are observed
concerning the number of opinions surviving in the (frozen) final
state. This can be a situation of {\em consensus}, in which all agents
share the same opinion, {\em polarization}, in which a finite number
of groups with different opinions survive, or {\em fragmentation},
with a final number of opinions scaling with the system size. 

In this Letter, we propose a model of opinion dynamics in which a
consensus-polarization-fragmentation non-equilibrium phase transition
is driven by external noise, intended as an `irresolute attitude' of
the agents in making decisions. The primary attribute of the model is
that it is based on a negotiation process, in which memory and
feedback play a central role. Moreover, apart from the consensus
state, no configuration is frozen: the stationary states with several
coexisting opinions are still dynamical, in the sense that the agents
are still able to evolve, in contrast to the Axelrod
model~\cite{axelrod}.

Let us consider a population of $N$ agents, each one endowed with a
memory, in which an a priori undefined number of opinions can be
stored. In the initial state, agents memories are empty.
At each time step, an ordered pair of neighboring agents is
randomly selected. This choice is consistent with the idea of {\em
directed attachment} in the socio-psychological literature (see for
instance~\cite{friedkin}). The
negotiation process is described by a local pairwise interaction rule:
a) the first agent selects randomly one of its opinions (or creates a
new opinion if its memory is empty) and conveys it to the second
agent; b) if the memory of the latter contains such an opinion, with
probability $\beta$ the two agents update their memories erasing all
opinions except the one involved in the interaction ({\em agreement}),
while with probability $1-\beta$ nothing happens; c) if the memory of
the second agent does not contain the uttered opinion, it adds such an
opinion to those already stored in its memory ({\em learning}).  Note
that, in the special case $\beta = 1$, the negotiation rule reduces to
the Naming Game rule~\cite{baronka}, a model used to describe the
emergence of a communication system or a set of linguistic conventions
in a population of individuals. In our modeling the parameter $\beta$
plays roughly the same role as the {\em probability of acknowledged
influence} in the socio-psychological
literature~\cite{friedkin}. Furthermore, as already stated for other
models~\cite{castellano}, when the system is embedded in heterogeneous
topologies, different pair selection criteria influence the
dynamics. In the {\em direct} strategy, the first agent is picked up
randomly in the population, and the second agent is randomly selected
among its neighbors. The opposite choice is called {\em reverse}
strategy; while the {\em neutral} strategy consists in randomly
choosing a link, assigning it an order with equal probability.\\ At
the beginning of the dynamics, a large number of opinions is created,
the total number of different opinions growing rapidly up to
$\mathcal{O}(N)$. Then, if $\beta$ is sufficiently large, the number
of opinions decreases until only one is left and the consensus state
is reached (as for the Naming Game in the case $\beta=1$). In the
opposite limit, when $\beta=0$, opinions are never eliminated,
therefore the only possible stationary state is the trivial state in
which every agent possesses all opinions. Thus, a non-equilibrium
phase transition is expected for some critical value $\beta_{c}$ of
the parameter $\beta$ governing the update efficiency.  In order to
find $\beta_{c}$, we exploit the following general stability
argument. Let us consider the consensus state, in which all agents
possess the same unique opinion, say $A$. Its stability may be tested
by considering a situation in which $A$ and another opinion, say $B$,
are present in the system: each agent can have either only opinion $A$
or $B$, or both ($AB$ state). The critical value $\beta_{c}$ is
provided by the threshold value at which the perturbed configuration
with these three possible states does not converge back to consensus.

\begin{figure}[t]
\centerline{
\includegraphics*[width=0.38\textwidth]{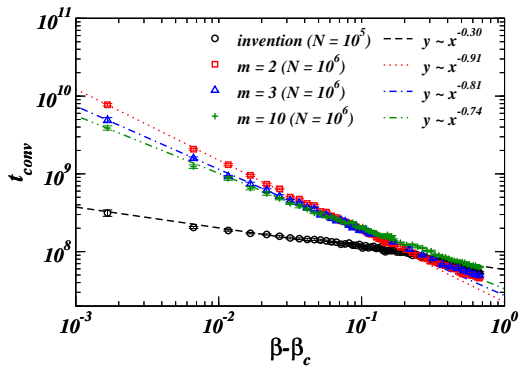}}
\caption{(Color online) Convergence time $t_{conv}$ of the model 
  as a function of
  $\beta-\beta_{c}$ (with $\beta_{c} = 1/3$) in the case of a fully-connected
  population of $N$ agents. We show data for the original model (circles) with
  unlimited number of opinions per agent, and for models with a finite number
  $m$ of different opinions. Increasing $m$, the power-law fits give exponents
  that differ considerably from the value $-1$.}
\label{fig1}
\end{figure}

\begin{figure}[t]
\centerline{
\includegraphics*[width=0.35\textwidth]{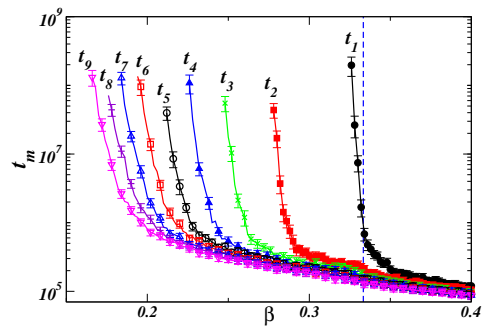}
}
\caption{(Color online) 
Time $t_{x}$ required to a population on a fully-connected
graph to reach a (fragmented) active stationary state with $x$
different opinions. For every $m > 2$, the time $t_{m}$ diverges at
some critical value $\beta_{c}(m) < \beta_{c}$.
}
\label{fig2}
\end{figure}

The simplest assumption in modeling a population of agents is the
homogeneous mixing (i.e. mean-field -- MF --
approximation), where the behavior of the system is completely described
by the following evolution equations for the densities $n_{i}$ of agents with
the opinion $i$:
\begin{eqnarray}
\nonumber d{n}_A/dt &= -  n_A n_B + 
\beta n_{AB}^2 +  \frac{3\beta -1}{2}n_A n_{AB}\\
d{n}_B/dt & = -  n_A n_B + \beta n_{AB}^2 + \frac{3\beta
-1}{2}n_B n_{AB},
\label{MF-neutro} 
\end{eqnarray}
and $n_{AB}=1-n_A-n_B$.  Imposing the steady state condition
$\dot{n}_A= \dot{n}_B=0$, we get three possible solutions: 1) $n_A=1$,
$n_B=0$, $n_{AB}=0$; 2) $n_A=0$, $n_B=1$, $n_{AB}=0$; and 3)
$n_A=n_B=b(\beta)$, $n_{AB}=1-2b(\beta)$ with $b(\beta)$=
$\frac{1+5\beta - \sqrt{1+10\beta+17\beta^2}}{4\beta}$ (and
$b(0)=0$). The study of the solutions' stability predicts a phase
transition at $\beta_{c}=1/3$. The maximum non-zero eigenvalue of the
linearized system around the consensus solution becomes indeed
positive for $\beta < 1/3$, i.e. the consensus becomes unstable, and
the population polarizes in the $n_{A}=n_{B}$ state, with a finite
density of undecided agents $n_{AB}$.  The model therefore displays a
first order non-equilibrium transition between the {\em frozen}
absorbing consensus state and an {\em active} polarized state, in
which global observables are stationary on average, but not frozen,
i.e. the population is split in three dynamically evolving parts (with
opinions $A$, $B$, and $AB$), whose densities fluctuate around the
average values $b(\beta)$ and $1-2b(\beta)$.

We have checked the predictions of Eqs.~(\ref{MF-neutro}) by numerical
simulations of $N$ agents interacting on a complete graph. Figure~\ref{fig1}
shows that the convergence time $t_{conv}$ required by the system to reach the
consensus state indeed diverges at $\beta_{c}=1/3$, with a power-law behavior
$(\beta-\beta_c)^{-a}$, $a\simeq 0.3$ \footnote{The low value of $a$, which
  moreover slightly decreases as the system size increases, does not allow to
  exclude a logarithmic divergence. This issue deserves more investigations
  that we leave for future work.}.  Very interestingly however, the analytical
and numerical analysis of Eqs.~(\ref{MF-neutro}) predicts that the relaxation
time diverges instead as $(\beta-\beta_c)^{-1}$. This apparent discrepancy
arises in fact because Eqs.~(\ref{MF-neutro}) consider that the agents have at
most two different opinions at the same time, while this number is unlimited
in the original model (and in fact diverges with $N$). Numerical simulations
reproducing the two opinions case allow to recover the behavior of $t_{conv}$
predicted from Eq.~(\ref{MF-neutro}) (see Fig.~\ref{fig1}).  We have also
investigated the case of a finite number $m$ of opinions available to the
agents. The analytical result $a=1$ holds also for $m=3$ (but analytical
analysis for larger $m$ becomes out of reach), whereas preliminary numerical
simulations performed for $m=3, 10$ with the largest reachable population size
($N=10^6$) lead to an exponent $a\simeq 0.74 \div 0.8$ (see
Fig.~\ref{fig1}).  More extensive and systematic simulations are in order to
determine the possible existence of a series of universality classes varying the
memory size for the agents. In any case, the models with finite
($m$ opinions) or unlimited memory define at least two clearly different
universality classes for this non-equilibrium phase transition between
consensus and polarized states (see~\cite{cris-grass} for similar findings in
the framework of non-equilibrium $q$-state systems).

Figure~\ref{fig2} moreover shows that the transition at
$\beta_{c}$ is only the first of a series of transitions: when
decreasing $\beta <\beta_{c}$, a system starting from empty initial
conditions self-organizes into a fragmented state with an increasing
number of opinions.  In principle, this can be shown analytically
considering the mean-field evolution equations for the partial
densities when $m> 2$ opinions are present, and studying, as a
function of $\beta$, the sign of the eigenvalues of a $(2^m -1) \times
(2^m -1)$ stability matrix for the stationary state with $m$
opinions. For increasing values of $m$, such a 
calculation becomes rapidly very demanding, thus we limit our analysis
to the numerical insights of Fig.~\ref{fig2}, from which we also get
that the number of residual opinions in the fragmented state follows
the exponential law $m(\beta) \propto \exp{[(\beta_{c} -\beta)/C]}$,
where $C$ is a constant depending on the initial conditions (not
shown).

\begin{figure}[t]
\centerline{ \includegraphics*[width=0.38\textwidth]{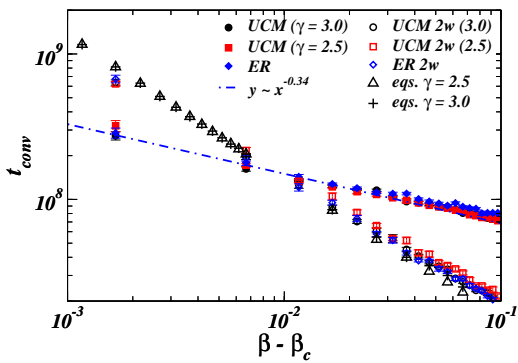} }
\caption{(Color online)
Convergence time $t_{conv}$ of the model as a function of
$\beta-\beta_{c}$ on networks with different topological properties:
the UCM networks with degree distributions $P(k)\sim k^{-\gamma}$,
$\gamma=2.5$ and $\gamma=3$, and the ER homogeneous random
graphs. Simulations are shown for networks of $N=10^5$ nodes
and average degree $\avk = 10$, both for $m=2$ ("2w", open symbols) and 
the original model with unlimited memory (filled symbols).
The numerical integration of Eqs.~\ref{NET-neutro}  
is in good agreement with the simulations.}
\label{fig3}
\end{figure}

We now extend our analysis to more general interactions topologies, 
in which agents are placed on the vertices of a network, and the edges define
the possible interaction patterns. When the network is a homogeneous
random one (Erd\"os-R\'enyi -- ER-- graph~\cite{erdos}), the degree
distribution is peaked around a typical value $\avk$, and the evolution
equations for the densities when only two opinions are present provide
the same transition value $\beta_{c}=1/3$ and the same exponent $-1$ for
the divergence of $t_{conv}$ as in MF. Figure~\ref{fig3} also shows that 
the exponent is also the MF one when the number of opinions is not limited.

Since any real negotiation process takes place on social groups, whose
topology is generally far from being homogeneous, we have simulated the model
on various uncorrelated heterogeneous networks (using the Uncorrelated
Configuration model --UCM-- model~\cite{UCM}), with power-law degree
distributions $P(k)\sim k^{-\gamma}$ with exponents $\gamma=2.5$ and
$\gamma=3$. 

Very interestingly, the model still presents a consensus-polarization
transition, in contrast with other opinion-dynamics models, like for instance
the Axelrod model~\cite{ax_no_trans}, for which the transition disappears for
heterogeneous networks in the thermodynamic limit. Moreover, Fig.~\ref{fig3}
reports the convergence time $t_{conv}$ vs. $(\beta-\beta_c)^{-a}$, showing
that at least two different universality classes are again present, one for
the case with a finite ($m=2$) number of opinions ($a = 1$) and one for the
case with unlimited memory ($a \simeq 0.3$). The exponents measured are in
each case compatible (up to the numerical precision) with the corresponding MF
exponents (see. Fig.~\ref{fig3}).

\begin{figure}[t]
\centerline{
\includegraphics*[width=0.35\textwidth]{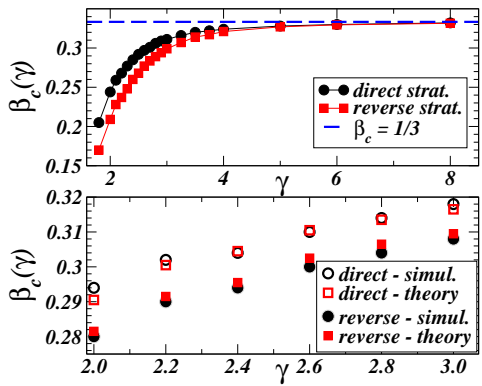} 
}
\caption{(Color online) Behavior of the critical
  value $\beta_{c}$ as a function of the exponent $\gamma$ of the degree
  distribution $P(k) \sim k^{-\gamma}$, as obtained from the numerical
  solution of the evolution equations for $n_i^k$, for both direct (black
  circles) and reverse (red squares) strategies. Bottom: comparison between
  the values of $\beta_{c}(\gamma)$ obtained from the equations and from
  numerical simulations on UCM networks of $N=10^3$ agents, for direct
  (open symbols) and reverse (full symbols) strategies.}
\label{fig4}
\end{figure}

To understand these numerical results, we analyze, as for the
fully connected case, the evolution equations for the case of two
possible opinions.  Such equations can be written for general
correlated complex networks whose topology is completely defined by
the degree distribution $P(k)$, i.e. the probability that a node has
degree $k$, and by the degree-degree conditional probability $P(k'|k)$
that a node of degree $k'$ is connected to a node of degree $k$
(Markovian networks).  Using partial densities $n_A^k=N_A^k/N_k$,
$n_B^k=N_B^k/N_k$ and $n_{AB}^k=N_{AB}^k/N_k$, i.e. the densities on
classes of degree $k$, one derives mean-field type equations in
analogy with epidemic models. Let us consider for definiteness the
neutral pair selection strategy, the equation for $n_{A}^{k}$
is in this case
\begin{eqnarray}
\nonumber
&&\frac{dn_A^k}{dt}= -\frac{k n_A^k}{\avk } \sum_{k'} P(k'|k) n_B^{k'}
-\frac{k n_A^k}{2 \avk}  \sum_{k'} P(k'|k) n_{AB}^{k'}+ \\
\label{NET-neutro}
&+&\frac{3\beta k n_{AB}^k}{2\avk}   \sum_{k'} P(k'|k) n_A^{k'} +
\frac{\beta k n_{AB}^k}{\avk}  \sum_{k'} P(k'|k) n_{AB}^{k'}~,
\end{eqnarray}
\noindent and similar equations hold for $n_{B}^{k}$ and $n_{AB}^{k}$.
The first term corresponds to the situation in which an agent of
degree $k'$ and opinion $B$ chooses as second actor an
agent of degree $k$ with opinion $A$.  The second term corresponds to
the case in which an agent of degree $k'$ with opinions
$A$ and $B$ chooses the opinion $B$, interacting with an agent of
degree $k$ and opinion $A$. The third term is the sum of two
contributions coming from the complementary interaction; while the
last term accounts for the increase of agents of degree $k$ and
opinion $A$ due to the interaction of pairs of agents with $AB$
opinion in which the first agent chooses the opinion $A$. \\ Let us
define $\Theta_i$ = $\sum_{k'} P(k'|k) n_i^{k'}$, for $i =A, B, AB$.
Under the uncorrelation hypothesis for the degrees of neighboring
nodes, i.e.  $P(k'|k)= k'P(k')/\avk$, we get the following relation
for the total densities $n_i=\sum_k P(k)n_i^k$,
\begin{equation}
\label{evolut-mixedNET}
\frac{d(n_A-n_B)}{dt}= \frac{3\beta-1}{2} \Theta_{AB} (\Theta_A -
\Theta_B).
\end{equation}
If we consider a small perturbation around the consensus state
$n_A=1$, with $n_A^k \gg n_B^k$ for all $k$, we can argue that
$\Theta_A-\Theta_B=\sum_k k P(k) (n_A^k -n_B^k)/\avk$ is still positive,
i.e.  the consensus state is stable only for $\beta > 1/3$.  In other
words, the transition point does not
change in heterogeneous topologies when the neutral strategy is
assumed. This is in agreement with our numerical simulations, and in 
contrast with the other selection strategies. Figure~\ref{fig4}
displays indeed
the values of the critical parameter $\beta_{c}(\gamma)$ as a function of
the exponent $\gamma$ as computed from the evolution equations of the
densities $n_i^k$ (that can be derived
similarly to Eqs.~(\ref{NET-neutro})), and as obtained from
numerical simulations. In such topologies, the phase transition is
shifted towards lower values of $\beta$, both for direct and
reverse strategies, revealing that a preferential bias in the
choice of the role played by hubs has a strong effect on the negotiation
process. Reducing the skewness of $P(k)$ (increasing $\gamma$), the
critical value of $\beta$ converges to $1/3$.

In conclusion, we have proposed a new model of opinion dynamics based
on agents negotiation in which instead memory and feedback are the
essential ingredients. We have shown that a non-trivial
consensus-polarization-fragmentation phase transition is observed in
terms of a control parameter representing the efficiency of the
negotiation process.  We have elucidated the mean-field dynamics, on
the fully connected graph as well as on homogeneous and heterogeneous
complex networks, using a simple continuous approach. We have shown
that the model presents a discontinuous phase transition between
consensus and polarized states featuring at least two different
universality classes, one for the case with $m=2$ opinions and one for
the case with an unlimited number of opinions.  In both cases we have
measured the critical exponent describing the divergence of the
convergence time and shown that they do not seem to depend on the
specific interaction topology. We argue that systems with any finite
number $m$ of opinions should fall in the $m=2$ class. Although this
point clearly deserves a deeper numerical investigation, we expect
that the behavior of the model with initial invention (unlimited
memory) may be due to the different spatial and temporal organization
of opinions in the inventories. It would also be interesting to study
the more realistic scenario in which the `irresolute attitude' of the
agents is modeled as a quenched disorder rather than a global external
parameter. {\em Acknowledgments} The authors wish to thank
V.D.P. Servedio for an important obervation concerning
eq.~(\ref{MF-neutro}).  A. Baronchelli and V.  L. are partially
supported by the EU under contract IST-1940 (ECAgents).  A.  Barrat
and L.D. are partially supported by the EU under contract 001907
(DELIS).

\end{document}